\documentclass[pdftex,twocolumn,epjc3]{svjour3}          

\RequirePackage[T1]{fontenc}

\smartqed  
\RequirePackage{amsmath}
\RequirePackage{amssymb}
\RequirePackage{graphicx}
\RequirePackage{mathptmx}      
\RequirePackage{flushend}
\RequirePackage[numbers,sort&compress]{natbib}
\RequirePackage[colorlinks,citecolor=blue,urlcolor=blue,linkcolor=blue]{hyperref}

\journalname{Eur. Phys. J. C}

\newcommand{\Fstar}{\raisebox{.2ex}{$\stackrel{*}{F}$}{}}

\begin{document}
\title{Born-Infeld magnetars: larger than classical toroidal magnetic fields and implications for gravitational-wave astronomy}
\titlerunning{Born-Infeld magnetars}
\author{Jonas P. Pereira\thanksref{e1,addr1,addr2} \and Jaziel G. Coelho\thanksref{e2,addr3,addr4} \and Rafael C. R. de Lima\thanksref{e3,addr5}
}                     

\thankstext{e1}{e-mail: jonas.pereira@ufabc.edu.br}
\thankstext{e2}{e-mail: jaziel.coelho@inpe.br}
\thankstext{e3}{e-mail: rafael.lima@udesc.br}

\institute{Universidade Federal do ABC, Centro de Ci\^encias Naturais e Humanas, Avenida dos Estados 5001, 09210-170, Santo Andr\'e, SP, Brazil\label{addr1} \and Mathematical Sciences and STAG Research Centre,
University of Southampton, Southampton, SO17 1BJ, United Kingdom\label{addr2} \and  Divis\~ao de Astrof\'isica, Instituto Nacional de Pesquisas Espaciais, Avenida dos Astronautas 1758, 12227-010, S\~ao Jos\'e dos Campos, SP, Brazil\label{addr3} \and Departamento de F\'isica, Universidade Tecnol\'ogica Federal do Paran\'a, 85884-000 Medianeira, PR, Brazil\label{addr4} \and Universidade do Estado de Santa Catarina, Av. Madre Benvenuta, 2007 Itacorubi, 88.035-901, Florian\'opolis, Brazil\label{addr5}
}
\date{Received: date / Revised version: date}
%
\maketitle

\begin{abstract}
Magnetars are neutron stars presenting bursts and
outbursts of X- and soft-gamma rays that can be understood with the presence of very large magnetic fields. Thus, nonlinear electrodynamics should be 
taken into account for a more accurate description of such compact systems. We study that in the context of ideal magnetohydrodynamics and make a realization of our analysis to the case of the well known Born-Infeld (BI) electromagnetism in order to come up with some of its astrophysical consequences. We focus here on toroidal magnetic fields as motivated by already known magnetars with low dipolar magnetic fields and their expected relevance in highly magnetized stars. We show that BI electrodynamics leads to larger toroidal magnetic fields when compared to Maxwell's electrodynamics. Hence, one should expect higher production of gravitational waves (GWs) and even more energetic giant flares from nonlinear stars. Given current constraints on BI's scale field, giant flare energetics and magnetic fields in magnetars, we also find that the maximum magnitude of magnetar ellipticities should be $10^{-6}-10^{-5}$. Besides, BI electrodynamics may lead to a maximum increase of order $10\%-20\%$ of the GW energy radiated from a magnetar when compared to Maxwell's, while much larger percentages may arise for other physically motivated scenarios. Thus, nonlinear theories of the electromagnetism might also be probed in the near future with the improvement of GW detectors.
\end{abstract}
%

\section{Introduction}
\label{int}

Soft Gamma Repeaters (SGRs) and Anomalous X-Ray pulsars (AXPs) (also known as magnetars~\citep[see, e.g.,][and references therein]{2014ApJS..212....6O}) are spectacular astrophysical systems which hold the unique possibility of probing yet unknown particle physics under extremely high magnetic fields. Believed to be ``lonely wolves'', among other properties such objects are observationally characterized by outbursts of X-ray and soft gamma-ray flares and have rotational periods $P\sim(2$--$12)$~s and slowing down rates $\dot{P}\sim(10^{-15}-10^{-10})$~s/s~\citep{2017ARA&A..55..261K}. Duncan and Thompson
\citep{1992ApJ...392L...9D,1995MNRAS.275..255T} proposed that they would be neutron stars (NSs) with huge magnetic fields (magnetars), of the order of $10^{14}-10^{15}$ G. One of the reasons for this would be their transient activities in the form of giant flares, whose typical luminosities are $10^{44}$-$10^{47}$ erg s$^{-1}$ \citep{2017ARA&A..55..261K}, not possible in general to be powered by their rotational energy.
\footnote{Recently, though, the authors of Ref. \cite{2017A&A...599A..87C} have discussed the possibility of some SGRs/AXPs being rotation-powered NSs and explored the entire range of NS parameters allowed by the conditions of stability of the star, not only fiducial parameters.} Notwithstanding, three counterexamples of magnetars with low surface magnetic fields are already known \citep{2010Sci...330..944R,2012ApJ...754...27R,2014ApJ...781L..17R}. 
This apparent ``glitch'' of the magnetar model has motivated different scenarios for the explanation of SGRs/AXPs, e.g.: the possibility of a fallback disk slowing down a neutron star pulsar up to its current spin period ~\citep[see, e.g.,][]{2011ApJ...732L...4A,2013ApJ...764...49T}; drift waves near the light-cylinder of NSs \citep[see][and references therein]{2010ARep...54..925M}; exotic scenarios involving quark stars~\citep{2006MNRAS.373L..85X}; and massive, fast rotating, highly magnetized white dwarfs to explain these types of sources~\citep{2012PASJ...64...56M,2014PASJ...66...14C,2017MNRAS.465.4434C}. Low-B values are related to surface poloidal magnetic fields, obtained by assuming that the spinning down of a star is due to its magnetic dipolar radiation \citep{1986bhwd.book.....S}, namely $B_{pol}\simeq 3\times 10^{19} (P\dot{P})^{1/2}$ G~\citep{2017ARA&A..55..261K}. Thus, low-B observations rendered the issue of magnetars much more complex than just outbursts and high dipolar magnetic fields and strengthened the relevance of toroidal magnetic fields in stars. 

Numerical simulations of ordinary twisted magnetic fields in magnetars suggest they should be poloidal field dominated, which is at odds with some inferences from observations~\citep[see details in][and references therein]{2010MNRAS.406.2540C,2013MNRAS.435L..43C}, such as the three above mentioned low-poloidal-B magnetars. This has been addressed in the context of a modified twisted magnetic field, where ratios of the toroidal energy to the total magnetic energy in stars could be up to $90\%$ \citep{2013MNRAS.435L..43C}. This has been done in the context of Maxwellian electromagnetism and toroidal fields as high as $10^{15}$ G could be obtained. However, for such large magnetic fields, nonlinearities of the electromagnetism should play a significant role in the observational properties of the systems bearing them. Motivations for nonlinear electrodynamics in the context of high fields naturally come from QED \citep{2010PhR...487....1R} but in this paper we focus on an older approach, due to Born and Infeld \citep{1934RSPSA.144..425B}, interesting even nowadays due to the reasons that follow.

Born-Infeld's electromagnetism was conceived with the purpose of healing the energy singularities of point-like charged particles such as the electron and is motivated by the finiteness of physical observables in special relativity \citep{1934RSPSA.144..425B}. It was showed in the 1980s that it is consistent with the low-energy limit of string theory and this has, since then, drawn very much attention to it  \citep[see][and references therein]{1997hep.th....2087R}. It already has an exact black hole solution \citep[see for instance][and references therein]{2015PhRvD..91f4048P} though in this context singularities are unavoidable \citep{2001PhRvD..63d4005B}. When applied to the hydrogen atom, Born-Infeld theory as the description for electromagnetic interactions meets the observational spectrum of this atom only if its scale factor $b$ is larger than the one inferred by Born and Infeld themselves within the unitarian viewpoint \citep{2006PhRvL..96c0402C,2011PhLA..375.1391F,2018EPJC...78..143A}, approximately $10^{15}$ statvolt cm$^{-1}$ (or also $10^{15}$~G) \citep{1934RSPSA.144..425B} (around 100 times QED's critical fields \cite{2010PhR...487....1R}). There are also constraints on $b$ coming from particle accelerators. It has been recently shown that LHC light-by-light scattering in Pb-Pb collisions would restrict $b$ to be larger than $(100\, GeV)^2\approx 10^{23}$ statvolt cm$^{-1}$ $\approx 10^ {23}$ G [$1 eV^2\approx 14.4$~G $\approx 14.4$ statvolt cm$^{-1}$] \citep{2017PhRvL.118z1802E}. However, it is important to point out that the kinematic cuts made in ATLAS and the self-consistency of the linear analysis used for cross section calculations are such that this experiment does not allow for the test of smaller values of $b$ 
\citep{2017PhRvL.118z1802E}. All the above means that the window $10^{15}$ G $\lesssim b\lesssim 10^{23}$~G cannot be assessed by LHC experiments so far and hence they leave a ``hole'' in the probe of the $b$ parameter. Given that magnetars are believed to have surface fields as high as $10^{15}$ G (even larger fields in their interiors), they could be the ideal systems for assessing scale fields to the Born-Infeld theory in the region the LHC cannot. This is important and is a motivation our analysis because the Born-Infeld electrodynamics can only be disregarded when its whole space of parameters is excluded. Besides, magnetars could also be ideal testbeds for other nonlinear theories of the electromagnetism. Impressively enough, perhaps due to the intrinsic difficulties already present in the Maxwell theory, such analysis seems scarce in the literature of magnetars. We try to partially fill this gap in here.

We structure this work in the following way. Next section is devoted to the deduction of the field equations in nonlinear electromagnetism for the case of ideal magnetohydrodynamics.
In Sec. \ref{BItoroidal} we work out a realization of nonlinear toroidal fields, by focusing on some consequences of it for the case of Born-Infeld electrodynamics. Some astrophysical consequences thereof are investigated in Sec. \ref{astro-implications}, especially related to the increase of the magnetic ellipticity and gravitational-wave energy budget when compared to their Maxwellian counterparts. Finally, discussions and conclusions are given in Sec. \ref{discussion}. We work with Gaussian and geometric units.  

\section{Nonlinear electrodynamics in ideal magnetohydrodynamics}

For a Lagrangian density $L$ depending upon both invariants of the electromagnetism, $F\doteq F_{\mu\nu}F^{\mu\nu}$ and $G\doteq F_{\mu\nu}\Fstar^{\mu\nu}$, $\Fstar^{\mu\nu}$ the dual of $F_{\mu\nu}$ \citep{1975ctf..book.....L}, the appropriate electromagnetic action is
\begin{equation}
S_{em}=a\int d^4x\, \sqrt[]{-g} L (F,G)- \int d^4x\, \sqrt[]{-g}j^{\mu}A_{\mu},\label{action-em-fields}
\end{equation}
where $a$ is a numerical factor that depends upon the system of units used, $g$ is the determinant of the spacetime metric $g_{\mu\nu}$ (assumed in this work to be given), $j^{\mu}$ is the current four-vector of the system and $A_{\mu}$ is the four-potential of the electromagnetic fields. We work with Gaussian units, which means we take $a=-1/16\pi$. By assuming that $F_{\mu\nu}=A_{\mu;\nu}-A_{\nu;\mu}=A_{\mu,\nu}-A_{\nu,\mu}$ and varying Eq. (\ref{action-em-fields}) with respect to $A_{\mu}$, we obtain the field equations
\begin{eqnarray}
\partial_{\mu} (\sqrt[]{-g}L_F F^{\mu\nu}+\sqrt{-g}L_G\Fstar^{\mu\nu})&=&4\pi \sqrt[]{-g} j^{\nu}\;\;\; \mbox{or} \nonumber \\  
(L_F F^{\mu\nu}+L_G \Fstar^{\mu\nu})_{;\mu}&=& 4\pi j^{\nu},\label{eq-NLED} 
\end{eqnarray}
complemented with (a natural consequence of the definition of $F_{\mu\nu}$) \citep{1975ctf..book.....L}
\begin{equation}
\partial_{\mu}(\sqrt[]{-g} \Fstar^{\mu\nu})=0\;\;\; \mbox{or}\;\;\; (\Fstar^{\mu\nu})_{;\mu}=0\label{eq-from-four-pot}
\end{equation}
where $L_{X}\doteq \partial L/\partial X$, $X=(F,G)$, $\Fstar^{\mu\nu} \doteq \eta^{\mu\nu\alpha\beta}
F_{\alpha\beta}/(2\sqrt{-g})$, $\eta^{0123}\doteq +1$, is a totally antisymmetric tensor. Note that since the Lagrangian must be an even function of the invariant $G$ (due to symmetry requirements from electrodynamics), $L_G$ is an odd function of $G$, which means it is zero when the fields are orthogonal. 

In order to simplify our description and evidence the physical picture involved in our analysis, we assume now the case where the electromagnetic fields are orthogonal and take the spacetime metric to be the Minkowski metric. Crossed fields naturally arise in the context of ideal magnetohydrodynamics (MHD) [see e.g., \citep{1986bhwd.book.....S,2009lema.book.....S}], which we will consider throughout this work, and the use of a flat spacetime metric, though being an idealized case, will allow us to find the maximal changes of physical quantities regarding Maxwell's electrodynamics.~\footnote{Indeed, when general relativistic corrections are taken into account, quantities such as the magnetic fields should decrease with respect to their flat spacetime counterparts by a function depending on the compactness factor \citep{2004MNRAS.352.1161R,2015ApJ...799...23B,2017A&A...599A..87C}.
We plan to investigate the case of nonlinear electrodynamics in curved geometries as due to magnetars in a forthcoming work.}
As commented previously, $L_G=0$ for the case under interest and Eqs. (\ref{eq-NLED}) and (\ref{eq-from-four-pot}) can be cast as (we restore Gaussian units here)
\begin{equation}
\nabla\cdot (L_F\vec{E})=4\pi \rho\label{divE},
\end{equation}
\begin{equation}
\nabla\cdot \vec{B}=0\label{divB},
\end{equation}
\begin{equation}
\nabla\times \vec{E}=-\frac{1}{c}\frac{\partial \vec{B}}{\partial t}\label{rotE}
\end{equation}
and
\begin{equation}
\nabla\times (L_F\vec{B})=\frac{1}{c}\frac{\partial L_F\vec{E}}{\partial t}+\frac{4\pi}{c} \vec{j}\label{rotB},
\end{equation}
where we have defined $\rho$ as the system's charge density and $\vec{j}$ its current vector.

Since $\rho$ and $\vec{j}$ are given aspects of a system, one could take as reference Maxwell's electromagnetism. There, $4\pi \rho= \nabla\cdot \vec{E}_{Ma}$ and $\frac{4\pi}{c} \vec{j}=\nabla\times (\vec{B}_{Ma}) -\frac{1}{c}\frac{\partial \vec{E}_{Ma}}{\partial t}$, which allows us to recast Eqs. (\ref{divE}) and (\ref{rotB}) as
\begin{equation}
\nabla\cdot (L_F\vec{E}-\vec{E}_{Ma})=0\label{divEwMa}
\end{equation}
and
\begin{equation}
\nabla\times (L_F\vec{B}-\vec{B}_{Ma})=\frac{1}{c}\frac{\partial}{\partial t}(\partial L_F\vec{E}-\vec{E}_{Ma})\label{rotBwMa}.
\end{equation}
Care must be taken at this point due to Eqs. (\ref{divB}) and (\ref{rotE}), which are formally identical to their Maxwellian counterparts but are related to the fields $\vec{E}$ and $\vec{B}$, instead of $\vec{B}_{Ma}$ and $\vec{E}_{Ma}$. This means, for instance, that a dipolar magnetic field in Maxwell's electromagnetism cannot in general have the same functional form in nonlinear electrodynamics.

Nonetheless, for a very good conductive region of a rigidly rotating star with an angular frequency $\vec{\omega}$ in the regime of small velocities $v/c\ll 1$ ($v\doteq ||\vec{v}||$) , one has that (ideal MHD \citep{1986bhwd.book.....S,2009lema.book.....S})
\begin{equation}
\vec{E}=-\frac{\vec{v}}{c}\times\vec{B}=-\frac{\vec{\omega}\times \vec{r}}{c}\times \vec{B}\label{solEconduct},
\end{equation}
which comes from the assumption that Ohm's law also holds for nonlinear electrodynamics, taken in this work as a first approach, and the consideration the current density is finite in the limit of infinite conductivity. \footnote{Let us elaborate on this point. If one assumes the validity of Ohm's law also in nonlinear electrodynamics, then in a locally comoving reference frame (or rest-frame) $K'$ (with respect to a conducting fluid), one has that $\vec{j}'=\sigma \vec{E}'$, where $\vec{j}'$ and  $\vec{E}'$ are the current density and the (nonlinear) electric field there, respectively, and $\sigma$ is the conductivity of the fluid. Take now the limit $\sigma \rightarrow \infty$ (ideal conductor). If one assumes that $\vec{j}'$ is finite (could be any), then, necessarily, $\vec{E}'=\vec{0}$. Make use now of another inertial coordinate system $K$ (usually called the laboratory frame) such that $K'$ moves with respect to it with velocity $\vec{v}$. When $v/c\ll 1$, it follows that $\vec{E}'=\vec{E}+\vec{v}\times \vec{B}/c$ \citep{1975ctf..book.....L}. (Due to the smallness of the charge density in a conductor after a small characteristic time \citep{2001imhd.book.....D}, from the transformation laws of four-vectors in the limit of $v/c\ll 1$, it also follows that $\vec{j}'=\vec{j}$.) Thus, in the limit of ideal MHD, Eq. (\ref{solEconduct}) follows as a kinematic constraint, and hence would be valid for any theory of the electromagnetism.}

If one assumes that the magnetic field lines are dragged along the motion of the star (frozen), then it follows that (Alfv\'en's theorem) \citep{2009lema.book.....S}
\begin{equation}
\frac{\partial \vec{B}}{\partial t}=\nabla \times (\vec{v}\times \vec{B})=\nabla \times (\vec{\omega}\times \vec{r}\times \vec{B})\label{frozen-induc}.
\end{equation}
However, in nonlinear electrodynamics this point might be subtler in principle. If one inserts $\vec{E} = \vec{j}/\sigma - \vec{v}\times \vec{B}/c$ (from Ohm's law in first order in $v/c$; see footnote) into Eq. (\ref{rotE}), takes $\sigma$ to be constant, assumes that $\partial L_F \vec{E}/\partial t$ is negligible when compared to the current density (which could be justified due to the fact that in a given limit the equations of nonlinear electrodynamics tend to the Maxwell ones where that holds true for large $\sigma$ \citep{2001imhd.book.....D}), one has that
\begin{equation}
\frac{\partial \vec{B}}{\partial t}= \nabla \times (\vec{v}\times \vec{B}) - \frac{c^2}{4\pi\sigma}\nabla\times \nabla \times (L_F\vec{B})\label{frozen_nonlinear}.
\end{equation}
One sees from the above equation that Eq. (\ref{frozen-induc}) is recovered when $\sigma \rightarrow \infty$ (ideal MHD) and that Eq. (\ref{solEconduct}) arises in this limit. The above means that the magnetic flux through any closed loop in a perfectly conducting medium is constant also in nonlinear electrodynamics if the conditions leading to Eq. (\ref{frozen_nonlinear}) hold. We leave analyses of other scenarios (for instance finite $\sigma$, violation of Ohm's law, etc.) to be investigated elsewhere.

From Eq. (\ref{divEwMa}), one has 
\begin{equation}
L_F\vec{E}=\vec{E}_{Ma}+\nabla \times \vec{C}\label{solE},
\end{equation}
where $\vec{C}$ is an arbitrary vector.
If one assumes that the time derivative of $\vec{C}$ is negligible [which should be justified by the assumption of large conductivity and would be equivalent to the disregard of the time derivative term of Eq.~(\ref{rotB}) w.r.t. $\vec{j}$, as happens in Maxwell's ideal magnetohydrodynamics \citep{2009lema.book.....S}], then it follows from Eq. (\ref{rotBwMa}) that
\begin{equation}
L_F\vec{B}=\vec{B}_{Ma} + \nabla f\label{solB},
\end{equation}
where $f$ is also an arbitrary function. One can solve Eq. (\ref{solB}) for $\vec{B}$ (given a nonlinear electrodynamics), which, after replaced in Eq. (\ref{divB}), will fix the function $f$. 

From Eqs. (\ref{solE}), (\ref{solB}) and (\ref{solEconduct}), it immediately follows that
\begin{equation}
\vec{E}_{Ma} + \nabla \times \vec{C}= -\frac{\vec{\omega}\times \vec{r}}{c}\times (\vec{B}_{Ma} + \nabla f),
\end{equation}
which from the Maxwellian relationship of the fields [$\vec{E}_{Ma}=-(\vec{\omega}\times \vec{r})\times \vec{B}_{Ma}/c$] implies
\begin{equation}
\nabla \times \vec{C}= -\frac{\vec{\omega}\times \vec{r}}{c}\times \nabla f\label{C-f-relationship}.
\end{equation}
Thus, any non-null $f$ induces a non-null $\vec{C}$ and this is done through the assumption of the system's large conductivity, exactly as suggested by previous intuitive arguments. We stress that Eqs. (\ref{solE}), (\ref{solB}) and (\ref{C-f-relationship}) are simple just because they relate fields in different theories. Without the solutions for $\vec{B}_{Ma}$ and $\nabla f$, which are not easy in general, they are just indicative.

The electromagnetic energy density in nonlinear electrodynamics can be easily obtained with the mixed 00 component of the electromagnetic energy momentum tensor, which to the case of Eq. (\ref{action-em-fields}) for orthogonal fields is given by \citep{2017PhRvD..95b5011M}
\begin{equation}
16\pi T_{\mu}^{\nu}= -4L_FF_{\mu\alpha}F_{\gamma\beta}\,g^{\nu\gamma}g^{\alpha\beta}+L\,\delta_{\mu}^{\nu}\label{tmunu},
\end{equation}
where $\delta^{\nu}_{\mu}$ is the Kronecker delta function~\citep{1975ctf..book.....L}. When the hypothesis of very conductive media is taken into account ($|\vec{E}|\ll |\vec{B}|$), Eq. (\ref{tmunu}) implies that
\begin{equation}
16 \pi T_0^0\approx L\label{t00}.
\end{equation}

\section{Nonlinear toroidal fields}
\label{toroidal}

In order to have some insights into the influence of nonlinear electrodynamics on magnetars and motivated by the expected dominance of azimuthal fields inside such systems \citep{2013MNRAS.435L..43C}, here we focus on the nonlinear field equations for purely toroidal fields with axial symmetry (thought of as a rough approximation to the real scenario), such that $\vec{B}_t=B_{\phi}(r,\theta)\hat{\phi}$, where $\hat{\phi}$ is the azimuthal unit vector. This field is such that Eq. (\ref{divB}) is automatically satisfied. From Eq.~(\ref{solEconduct}), the case of toroidal fields in rigidly rotating stars {(we define the $z$-axis such that $\vec{\omega}=\omega \hat z$, which means $\vec{v}=v\hat \phi$, $v$ being any)} result in $\vec{E}=\vec{0}$, {which automatically satisfies Eq. (\ref{rotE}) [as well as Eq. (\ref{frozen-induc})]} and from Eqs.~(\ref{solE}) and (\ref{C-f-relationship}) implies $\nabla f = \vec{0}$. Thus, Eqs. (\ref{solB}) (the only remaining nonlinear equation) leads to
\begin{equation}
L_FB_{\phi}= B^{Ma}_{\phi}\label{Btoroidal}.
\end{equation}

Since in the small field regime of nonlinear electrodynamics $L_F=1-|\mbox{something small}|$ \citep{2016ApJ...823...97C}, it follows generically from the above equation that the magnitudes of nonlinear toroidal fields in ideal MHD are in general larger than their Maxwellian counterparts. Given that, we investigate an interesting case with an exact solution in the next subsection. 

\subsection{Born-Infeld toroidal field analysis}
\label{BItoroidal}

Born-Infeld's Lagrangian is defined as  \citep[see for instance][]{2014PhRvA..89d3822D,2017PhRvD..95b5011M} 
\begin{equation}
L_{B.I}\doteq 4b^2\left(\sqrt{1+\frac{F}{2b^2}-\frac{G^2}{16b^4}}-1\right)\label{LB-I},
\end{equation}
where $b$ is the scale field of the theory and from the definition of the invariants of the electromagnetism, $F= 2(\vec{B}^2- \vec{E}^2)$ and $G^2=16(\vec{E}\cdot \vec{B})^2$. Thus, for orthogonal fields, our main interest in this work, the only relevant derivative to the Lagrangian is
\begin{eqnarray}
L_F=\frac{b}{\sqrt{b^2+\vec{B}^2- \vec{E}^2}}\approx \frac{b}{\sqrt{b^2+\vec{B}^2
}}\label{LFB-I},
\end{eqnarray}
since for very good conducting stars it follows that $|\vec{E}|\ll~|\vec{B}|$. To the case of Born-Infeld toroidal magnetic fields, Eqs. (\ref{Btoroidal}) and (\ref{LFB-I}) lead to a very simple solution, namely,
\begin{equation}
{B}_{\phi}=\frac{b{B}^{Ma}_{\phi}}{\sqrt[]{b^2-(B^{Ma}_{\phi})^2}}\label{BBItoroidal},
\end{equation}
which clearly evidences the increase nonlinear electrodynamics impinges on $B_{\phi}$.

For toroidal-dominated fields, it follows from Eqs. (\ref{t00}), (\ref{LB-I}) and (\ref{BBItoroidal}) that the energy densities in Born-Infeld and Maxwell's theories are related by
\begin{equation}
\frac{(T_0^0)_{BI}}{(T^0_0)_{Ma}}= 2\left(\frac{b}{B^{Ma}_{\phi}} \right)^2 \left\{\left[1-\left(\frac{B_{\phi}^{Ma}}{b} \right)^2\right]^{-\frac{1}{2}} -1 \right\},\label{energy_ratio_BI}
\end{equation}
which is always larger than the unit for $B_{\phi}^{Ma}<b$, needed for the consistency of Born-Infeld toroidal fields. This means that toroidal fields in nonlinear electrodynamics imply larger energy reservoirs when compared to Maxwell's predictions. 

\subsection{Ellipticities for Born-Infeld toroidal fields} 

Strong toroidal magnetic fields also have important implications for the ellipticity, as we discuss now. Assume  that the magnetic ellipticity of a star is proportional to the mean value of $B_{\phi}^2$, i.e.,
\begin{equation}
\epsilon=c_1 \langle B_{\phi}^2\rangle \label{ellpiticity},
\end{equation}
where $c_1$ is a constant that depends upon aspects of the star such as its radius, compactness, etc. Indeed, when toroidal magnetic fields are dominant, one has that \citep{1969ApJ...157.1395O}
\begin{equation}
\epsilon\propto -\frac{1}{E_g}\int dV B_{\phi}^2,
\end{equation}
where $E_g\propto M^2/R$ is the magnitude of the gravitational energy of the star and $dV$ is a volume element. Thus, by defining $\langle B_{\phi}^2\rangle \doteq \int dV B_{\phi}^2/V$, Eq. (\ref{ellpiticity}) ensues with $c_1$ depending on the compactness of the star and its radius, all byproducts of its microphysics.

From Eq. (\ref{BBItoroidal}), it follows that the Born-Infeld to the Max-wellian ellipticity ratio is
\begin{equation}
\frac{\epsilon_{BI}}{\epsilon_{Ma}}= \left(\frac{\dot{E}_{GW}^{BI}}{\dot{E}_{GW}^{Ma}}\right)^{\frac{1}{2}}\approx \left(\frac{\Delta{E}_{GW}^{BI}}{\Delta{E}_{GW}^{Ma}}\right)^{\frac{1}{2}}\approx \left [ 1-\left(\frac{B_{\phi}^{Ma}}{b} \right)^2 \right]^{-1}\label{GW},
\end{equation}
where $\dot{E}_{GW}$ is the energy loss due to gravitational waves. We have assumed that $B_{\phi}^{Ma}$ is slowly varying within the star (where it takes place), which might be taken for maximal estimates. Since $B_{\phi}^{Ma}<b$ from consistency of Eq. (\ref{BBItoroidal}), one has from the above equation that the production of gravitational waves is larger in the Born-Infeld theory than in Maxwell's. 

Assume now that $c_1$ is given, so as an upper limit to the norm of the magnetic ellipticity of a star and the Maxwellian toroidal magnetic field. (One could easily \textit{estimate} $B_{\phi}^{Ma}$ by means of magnetar's observables such as flare luminosities and upper limits to the ellipticities could be inferred from gravitational-wave analysis.) In this case, from Eq. (\ref{ellpiticity}), it follows that Maxwell's ellipticity is known. If one takes the true theory of the electromagnetism as Born-Infeld's, thus
\begin{equation}
b\geq B_{\phi}^{Ma}\left ( 1- \frac{1}{{\cal C}} \right)^{-\frac{1}{2}},\;\; {\cal C}\doteq \frac{|\epsilon_{ul}|}{|\epsilon_{Ma}|}
 \label{b-limit},
\end{equation}
where $\epsilon_{ul}$ stands for the observational upper limit of the magnetic ellipticity. From Eqs. (\ref{BBItoroidal}) and (\ref{ellpiticity}) one sees that when the measured ellipticity is the one connected with the Born-Infeld theory, then it follows that ${\cal C}>1$ automatically.
Besides, Eq. (\ref{b-limit}) is totally equivalent to the consistency of Eq. (\ref{BBItoroidal}) since the minimum value of $b$ is $B_{\phi}^{Ma}(1-1/{\cal C})^{-1/2}$, which is larger than $B_{\phi}^{Ma}$. Therefore, Born-Infeld electrodynamics is self-consistent. 

\section{Possible Astrophysical Implications of Born-Infeld magnetars}
\label{astro-implications}

LIGO measurements already constrain ellipticities in ordinary \textit{pulsars} and their magnitude should be smaller than approximately $10^{-6}$ \citep{2017ApJ...839...12A}.  The values of the ellipticities they found should be taken as upper limits since no continuous gravitational-wave signals have been detected from neutron stars. In what follows we do not assume the above value holds true for magnetars, but we rather \textit{find} an upper limit to it by means of outcomes of Born-Infeld theory applied to the hydrogen atom. We will focus on the systems supposed to have dominant toroidal fields such that the analysis of the previous section could be taken into account. This is actually believed to be the case in all magnetars and even possibly in several pulsars \citep{2013MNRAS.435L..43C}.

In this context, mean toroidal Maxwellian fields could be \textit{estimated} with giant flare events by dint of
\begin{equation}
(B_{\phi}^{Ma})^2= \frac{6 E_{fl}}{R^3}\label{Bphi-mean},
\end{equation}
where $R$ is the radius of the magnetar and $E_{fl}$ is the total energy released in the giant flare. Precise $|c_1|$ calculations for the case of a homogeneous ellipsoid show that \citep{1969ApJ...157.1395O}
\begin{equation}
|c_1|\approx \frac{R^4}{GM^2}\label{c1-precise},
\end{equation}
where $G$ is Newton's constant and $M$ is the magnetar's mass. From Eq. (\ref{b-limit}) [or Eq. (\ref{GW})], $|\epsilon_{ul}|$ could be constrained if a minimum value for $b$, $b_{min}$, was given, implying that
\begin{equation}
|\epsilon_{ul}|=\left[1-\left(\frac{B_{\phi}^{Ma}}{b_{min}}\right)^2 \right]^{-1}|\epsilon_{Ma}|.\label{epsilonulbmin}
\end{equation}

Maximum values for $B_{\phi}^{Ma}$ are estimated by taking the maximum energy released during giant flares, around $10^{47}$ erg to the magnetar SGR 1806--20 \citep{2015RPPh...78k6901T,2017ARA&A..55..261K}. In this case, toroidal magnetic fields of order $7.7\times 10^{14}$ G arise and from Eqs.~ (\ref{ellpiticity}) and (\ref{c1-precise}), by making use of fiducial magnetar parameters [$R=10$ km and $M=1.4$ M$_{\bigodot}$], it follows that  
$|\epsilon_{Ma}|= 1.20\times 10^{-6}$. Taking into account hydrogen experiment outcomes \citep{2018EPJC...78..143A}, one learns that the absolute minimum value for Born-Infeld's scale field is $b_{min}\approx 3.96 \times 10^{15}$ statvolt cm$^{-1}$ (or $b_{min}\approx 3.96 \times 10^{15}$ G) \citep{1934RSPSA.144..425B} [most recent data for the mass and charge of the electron have been used]. Thus, from the above and Eqs. (\ref{epsilonulbmin}) and (\ref{GW}), we have that 
$|\epsilon_{ul}|\approx 1.24\times 10^{-6}$ and $\Delta{E}^{BI}_{GW}/\Delta{E}^{Ma}_{GW}\lesssim 1.08$. 

Naturally, if larger values of $E_{fl}$ are the case (for instance associated with uncertainties in distance measurements or even newly-born magnetars), then larger ellipticities would emerge. If $E_{fl}=10^{47}-10^{48}$ erg, say $E_{fl}=5\times 10^{47}$ erg, which might be possible to giant flare events when they are associated with nonlinear electrodynamics [which increases the magnetic energy reservoir of highly magnetized stars, see Eq.~(\ref{energy_ratio_BI})], or are associated with smaller than usual short GRBs~\footnote{The authors of Ref. \cite{2005Natur.434.1098H} have estimated that an appreciable quantity of short GRBs might be related to extragalactic unstable magnetars. However, Ref. \cite{2017arXiv171106593T} has cast doubt on whether or not the recent subluminous short gamma-ray burst GRB 170817 A \citep{2017ApJ...848L..13A}, associated with GW 170817 \citep{2017PhRvL.119p1101A}, might be related to a magnetar flare.} [typical isotropic energy of ordinary short-GRBs are in the range $10^{49}-10^{52}$ erg \citep{2015JHEAp...7...73D,2017arXiv171106593T}], then $|\epsilon_{ul}|\approx 7.4\times 10^{-6}$ [$B_{\phi}^{Ma}=1.7\times 10^{15}$ G] and $\Delta{E}^{BI}_{GW}/\Delta{E}^{Ma}_{GW}\lesssim 1.53$. 
If, instead, only magnetic fields as high as $10^{15}$ G are possible in magnetars, from Eq. (\ref{GW}) one has that $\Delta{E}^{BI}_{GW}/\Delta{E}^{Ma}_{GW}\lesssim 1.14$, which in turn would imply in $|\epsilon_{ul}|\approx 2\times 10^{-6}$. Thus, upper limits to the magnitude of magnetic ellipticities of magnetars would be around $10^{-6}$ ($10^{-5}$) and the maximum increase of gravitational-wave energy released in Born-Infeld theory with respect to Maxwell's theory would be around $15\%$ (even larger than $50\%$) for flare energy up to $10^{47}$ ($10^{48}$) erg. 

\section{Discussions and conclusions}
\label{discussion}

In order to explain magnetars' observational properties, it seems accepted today in the literature that very large resultant magnetic fields should take place in the vicinities of their surfaces and also interiors. Fields as high as $10^{15}$~G should be present in order to function as energy reservoirs to the flare activities magnetars display. With such high fields, it seems reasonable to assume nonlinear electrodynamics could give a more precise description of magnetars and concomitantly they could be natural candidates for \textit{probing} nonlinear electrodynamics astrophysically. Given that there are known magnetars with low-poloidal magnetic fields, toroidal fields should be particularly relevant for highly magnetized stars. Fields of order $10^{15}$~G are a natural scale for magnetic fields in magnetars and this value is of the same order as the scale field in Born-Infeld's original theory. This was one of our motivations to our analysis.

Axially symmetric toroidal nonlinear fields have a very simple solution in the context of magnetars' ideal magnetohydrodynamics because electric fields are not induced. Note this would be the case only where toroidal fields would exist and be much larger than poloidal ones, believed to be the case just in the interior of magnetars \citep{2013MNRAS.435L..43C}. Besides, nonlinear electrodynamics leads to toroidal magnetic field increase when compared to their Maxwellian counterparts. This would imply the increase of the magnetic energy of a magnetar, exactly as we have showed, which could lead to more energetic flare activities. Besides, more magnetic energy could be converted into gravitational-wave energy, as also explicitly showed. The kinematic reason for so is the increase in magnitude of the Born-Infeld magnetic ellipticity when compared to its Maxwellian counterpart.

Hydrogen atom byproducts show that Born-Infeld's scale field should be larger than the one obtained by Born and Infeld themselves. Actually, such  a value  defines an absolutely lower limit to the scale field and has to be taken into account for physically relevant constraints. (For the upper limit on $b$ which could not have been assessed by the ATLAS experiment, the changes introduced by the Born-Infeld Lagrangian are negligible when compared to the Maxwell theory.) When that is done, Born-Infeld theory predicts an upper limit to the magnetic ellipticity of a magnetar, which also lead to an upper limit to the gravitational-wave energy it may emit. Giant flares are important in this case because they result in the largest energy budgets of the system, which according to the magnetar theory should have exclusive magnetic origins. We have showed that the upper limit to the magnitude of the ellipticity should be within the range $10^{-5}-10^{-6}$ for current observations of the magnetar SRG 1806-20. Born-Infeld's gravitational-wave energy for giant flare events of around $10^{47}$ erg would be at most $10\%-20\%$ larger than their classical counterparts, while for flare energy up to $10^{48}$ erg the maximum percentage could be much higher than $50\%$. 
Naturally, the previous incredibly large Born-Infeld gravitational-wave energy increase should be seen only as an indicative value, given that $b$ should be larger than $b_{min}$. When LIGO/VIRGO are able to constrain magnetar ellipticities, we see from Eq. (\ref{b-limit}) that minimum values for the Born-Infeld's scale field could be inferred astrophysically. This seems very interesting since it would work as a possible cross-check to the already available constraints on $b$ and be able to probe the region of parameters current experiments cannot.

It is argued that a fiducial upper limit to the norm of the ellipticity should be around $10^{-6}$ when asymmetries supported by anisotropic stresses
built up during the crystallization period of the crust are taken into account~\citep{2008PhLB..668....1K}. It is worth mentioning that Refs. \cite{2016ApJ...831...35D,2017EPJC...77..350D} have predicted GW amplitudes for all known \textit{pulsars} and, when an extremely optimistic case is considered, ellipticities should be at most $10^{-5}$ (for PSR J1846-0258). Thus, since the predicted GW amplitudes are extremely small, observation times of thousands of years would be needed even for advanced detectors such as aLIGO and AdVirgo, and the planned Einstein Telescope might not be able to detect these pulsars. All of the above actually evidences the relevance of finding the maximum increase of gravitational-wave energy in Born-Infeld theory with respect to Maxwell's theory. The conclusion in the context of Born-Infeld theory is that magnetars with giant flare energy $\gtrsim 10^{47}$ erg would be the most promising candidates for gravitational-wave detections and by consequence potential tests of nonlinear electrodynamics.

In a realistic stellar model, magnetic fields should present both poloidal and toroidal components and should be dependent upon the azimuthal coordinate and time too (besides $r$ and $\theta$). However, for slowly rotating highly magnetized nonlinear stars it seems reasonable to start with simplified stationary and axially symmetric models, as suggested by some MHD simulations \citep{2013MNRAS.435L..43C}. Therefore, one could conceive models in the form $\vec{B}=\vec{B}_{pol}(r,\theta) + \vec{B}_{tor}(r,\theta)$. Notwithstanding, in the context of nonlinear theories, finding $\vec{B}_{pol}$ and $\vec{B}_{tor}$ is expected to be even harder than in Maxwell's electrodynamics due to the natural coupling of these components. When toroidal fields are very large, though, one could uncouple the system of equations. Besides, it is also pending stability analysis for magnetic fields in the context of nonlinear electrodynamics and when Ohm's law does not hold. It is already known that Maxwell's theory in the context of ideal MHD leads to instabilities of purely poloidal or toroidal fields (see e.g., \cite{2011MNRAS.412.1394L,2011MNRAS.412.1730L} and references therein).
These analyses are interesting on their own and are left for future works.

Additionally, for describing more realistically nonlinear electrodynamics in magnetars and the effects thereof, curved geometries should also be taken into account. However, in a first approach, it seems justifiable to neglect them such that maximal field strength (and derived quantities) and a better understanding of the physics taking place in the above scenario could be obtained. This would allow one to find which aspects are more important to be focused on in more precise analyses. We have seen that magnetic ellipticities are clear candidates. Curved geometries may significantly affect them by changing the spatial distributions of magnetic fields. Besides, the presence of both poloidal and toroidal magnetic fields and their couplings might also cancel out one another effects on the magnetic ellipticity. Due to the connections of the above effects with magnetar GWs, we plan to investigate these issues elsewhere.

Summing up, we have showed that when ideal magnetohydrodynamics and nonlinear electrodynamics are taken into account in magnetars, axially symmetric toroidal fields should be larger than their Maxwellian counterparts. This implies larger magnetic energy and ellipticities, which could thus increase the emission of gravitational waves from nonlinear highly magnetized stars. Current constraints on the Born-Infeld theory, giant-flare energetics and magnetic fields in magnetars point to a maximum increase of $10\%-20\%$ of the energy emitted in the form of gravitational waves by Born-Infeld magnetars when compared to Maxwellian ones. When larger giant-flare energy are taken into account, in principle plausible due to the nonlinearities of the electromagnetism which could increase the magnetic energy reservoirs of magnetars, much higher percentages may appear. Thus, the possibility may arise for also probing nonlinear electrodynamics with the advancement of gravitational-wave detectors.

\begin{acknowledgements}
J.P.P. acknowledges the financial support given by Funda\c c\~ao de Amparo \`a Pesquisa do Estado de S\~ao Paulo (FAPESP) under grants No. 2015/04174-9 and 2017/21384-2. J.G.C. is likewise grateful to the support of FAPESP (2013/15088-0 and 2013/26258-4). R.C.R.L. acknowledges the support of Funda\c c\~ao de Amparo \`a Pesquisa e Inova\c c\~ ao do Estado de Santa Catarina (FAPESC) under grant No. 2017TR1761.
\end{acknowledgements}




\bibliographystyle{spphys}
\bibliography{ref}

\begin{thebibliography}{10}
\providecommand{\url}[1]{{#1}}
\providecommand{\urlprefix}{URL }
\expandafter\ifx\csname urlstyle\endcsname\relax
  \providecommand{\doi}[1]{DOI \discretionary{}{}{}#1}\else
  \providecommand{\doi}{DOI \discretionary{}{}{}\begingroup
  \urlstyle{rm}\Url}\fi

\bibitem{2014ApJS..212....6O}
S.A. {Olausen}, V.M. {Kaspi}, \apjs \textbf{212}, 6 (2014).
\newblock \doi{10.1088/0067-0049/212/1/6}

\bibitem{2017ARA&A..55..261K}
V.M. {Kaspi}, A.M. {Beloborodov}, Ann. Rev. Astron. Astrophys. \textbf{55}, 261
  (2017).
\newblock \doi{10.1146/annurev-astro-081915-023329}

\bibitem{1992ApJ...392L...9D}
R.C. {Duncan}, C.~{Thompson}, \apjl \textbf{392}, L9 (1992).
\newblock \doi{10.1086/186413}

\bibitem{1995MNRAS.275..255T}
C.~{Thompson}, R.C. {Duncan}, \mnras \textbf{275}, 255 (1995).
\newblock \doi{10.1093/mnras/275.2.255}

\bibitem{2017A&A...599A..87C}
J.G. {Coelho}, D.L. {C{\'a}ceres}, R.C.R. {de Lima}, M.~{Malheiro}, J.A.
  {Rueda}, R.~{Ruffini}, \aap \textbf{599}, A87 (2017).
\newblock \doi{10.1051/0004-6361/201629521}

\bibitem{2010Sci...330..944R}
N.~{Rea}, P.~{Esposito}, R.~{Turolla}, G.L. {Israel}, S.~{Zane}, L.~{Stella},
  S.~{Mereghetti}, A.~{Tiengo}, D.~{G{\"o}tz}, E.~{G{\"o}{\u g}{\"u}{\c s}},
  C.~{Kouveliotou}, Science \textbf{330}, 944 (2010).
\newblock \doi{10.1126/science.1196088}

\bibitem{2012ApJ...754...27R}
N.~{Rea}, G.L. {Israel}, P.~{Esposito}, J.A. {Pons}, A.~{Camero-Arranz}, R.P.
  {Mignani}, R.~{Turolla}, S.~{Zane}, M.~{Burgay}, A.~{Possenti}, S.~{Campana},
  T.~{Enoto}, N.~{Gehrels}, E.~{G{\"o}{\v g}{\"u}{\c s}}, D.~{G{\"o}tz},
  C.~{Kouveliotou}, K.~{Makishima}, S.~{Mereghetti}, S.R. {Oates}, D.M.
  {Palmer}, R.~{Perna}, L.~{Stella}, A.~{Tiengo}, \apj \textbf{754}, 27 (2012).
\newblock \doi{10.1088/0004-637X/754/1/27}

\bibitem{2014ApJ...781L..17R}
N.~{Rea}, D.~{Vigan{\`o}}, G.L. {Israel}, J.A. {Pons}, D.F. {Torres}, \apjl
  \textbf{781}, L17 (2014).
\newblock \doi{10.1088/2041-8205/781/1/L17}

\bibitem{2011ApJ...732L...4A}
M.A. {Alpar}, {\"U}.~{Ertan}, {\c S}.~{{\c C}al{\i}{\c s}kan}, \apjl
  \textbf{732}, L4 (2011).
\newblock \doi{10.1088/2041-8205/732/1/L4}

\bibitem{2013ApJ...764...49T}
J.E. {Tr{\"u}mper}, K.~{Dennerl}, N.D. {Kylafis}, {\"U}.~{Ertan}, A.~{Zezas},
  \apj \textbf{764}, 49 (2013).
\newblock \doi{10.1088/0004-637X/764/1/49}

\bibitem{2010ARep...54..925M}
I.F. {Malov}, Astronomy Reports \textbf{54}, 925 (2010).
\newblock \doi{10.1134/S1063772910100057}

\bibitem{2006MNRAS.373L..85X}
R.X. {Xu}, D.J. {Tao}, Y.~{Yang}, \mnras \textbf{373}, L85 (2006).
\newblock \doi{10.1111/j.1745-3933.2006.00248.x}

\bibitem{2012PASJ...64...56M}
M.~{Malheiro}, J.A. {Rueda}, R.~{Ruffini}, \pasj \textbf{64}, 56 (2012)

\bibitem{2014PASJ...66...14C}
J.G. {Coelho}, M.~{Malheiro}, \pasj \textbf{66}, 14 (2014).
\newblock \doi{10.1093/pasj/pst014}

\bibitem{2017MNRAS.465.4434C}
D.L. {C{\'a}ceres}, S.M. {de Carvalho}, J.G. {Coelho}, R.C.R. {de Lima}, J.A.
  {Rueda}, \mnras \textbf{465}, 4434 (2017).
\newblock \doi{10.1093/mnras/stw3047}

\bibitem{1986bhwd.book.....S}
S.L. {Shapiro}, S.A. {Teukolsky}, \emph{{Black Holes, White Dwarfs and Neutron
  Stars: The Physics of Compact Objects}} (Wiley-VCH, Weinheim, 1986)

\bibitem{2010MNRAS.406.2540C}
R.~{Ciolfi}, V.~{Ferrari}, L.~{Gualtieri}, \mnras \textbf{406}, 2540 (2010).
\newblock \doi{10.1111/j.1365-2966.2010.16847.x}

\bibitem{2013MNRAS.435L..43C}
R.~{Ciolfi}, L.~{Rezzolla}, \mnras \textbf{435}, L43 (2013).
\newblock \doi{10.1093/mnrasl/slt092}

\bibitem{2010PhR...487....1R}
R.~{Ruffini}, G.~{Vereshchagin}, S.~{Xue}, Phys. Rep. \textbf{487}, 1 (2010).
\newblock \doi{10.1016/j.physrep.2009.10.004}

\bibitem{1934RSPSA.144..425B}
M.~{Born}, L.~{Infeld}, Royal Society of London Proceedings Series A
  \textbf{144}, 425 (1934).
\newblock \doi{10.1098/rspa.1934.0059}

\bibitem{1997hep.th....2087R}
D.A. {Rasheed}, ArXiv High Energy Physics - Theory e-prints  (1997)

\bibitem{2015PhRvD..91f4048P}
J.P. {Pereira}, J.A. {Rueda}, \prd \textbf{91}(6), 064048 (2015).
\newblock \doi{10.1103/PhysRevD.91.064048}

\bibitem{2001PhRvD..63d4005B}
K.A. {Bronnikov}, \prd \textbf{63}(4), 044005 (2001).
\newblock \doi{10.1103/PhysRevD.63.044005}

\bibitem{2006PhRvL..96c0402C}
H.~{Carley}, M.K.H. {Kiessling}, Physical Review Letters \textbf{96}(3), 030402
  (2006).
\newblock \doi{10.1103/PhysRevLett.96.030402}

\bibitem{2011PhLA..375.1391F}
J.~{Franklin}, T.~{Garon}, Physics Letters A \textbf{375}, 1391 (2011).
\newblock \doi{10.1016/j.physleta.2011.02.012}

\bibitem{2018EPJC...78..143A}
P.N. {Akmansoy}, L.G. {Medeiros}, European Physical Journal C \textbf{78}, 143
  (2018).
\newblock \doi{10.1140/epjc/s10052-018-5643-1}

\bibitem{2017PhRvL.118z1802E}
J.~{Ellis}, N.E. {Mavromatos}, T.~{You}, Physical Review Letters
  \textbf{118}(26), 261802 (2017).
\newblock \doi{10.1103/PhysRevLett.118.261802}

\bibitem{1975ctf..book.....L}
L.D. {Landau}, E.M. {Lifshitz}, \emph{{The classical theory of fields}} (1975)

\bibitem{2009lema.book.....S}
D.D. {Schnack}, \emph{{Lectures in Magnetohydrodynamics}} (Springer, Berlin,
  2009)

\bibitem{2004MNRAS.352.1161R}
L.~{Rezzolla}, B.J. {Ahmedov}, \mnras \textbf{352}, 1161 (2004).
\newblock \doi{10.1111/j.1365-2966.2004.08006.x}

\bibitem{2015ApJ...799...23B}
R.~{Belvedere}, J.A. {Rueda}, R.~{Ruffini}, \apj \textbf{799}, 23 (2015).
\newblock \doi{10.1088/0004-637X/799/1/23}

\bibitem{2001imhd.book.....D}
P.A. {Davidson}, \emph{{An introduction to magnetohydrodynamics}} (2001)

\bibitem{2017PhRvD..95b5011M}
H.J. {Mosquera Cuesta}, G.~{Lambiase}, J.P. {Pereira}, \prd \textbf{95}(2),
  025011 (2017).
\newblock \doi{10.1103/PhysRevD.95.025011}

\bibitem{2016ApJ...823...97C}
J.G. {Coelho}, J.P. {Pereira}, J.C.N. {de Araujo}, \apj \textbf{823}, 97
  (2016).
\newblock \doi{10.3847/0004-637X/823/2/97}

\bibitem{2014PhRvA..89d3822D}
V.A. {De Lorenci}, J.P. {Pereira}, \pra \textbf{89}(4), 043822 (2014).
\newblock \doi{10.1103/PhysRevA.89.043822}

\bibitem{1969ApJ...157.1395O}
J.P. {Ostriker}, J.E. {Gunn}, \apj \textbf{157}, 1395 (1969).
\newblock \doi{10.1086/150160}

\bibitem{2017ApJ...839...12A}
B.P. {Abbott}, R.~{Abbott}, T.D. {Abbott}, M.R. {Abernathy}, F.~{Acernese},
  K.~{Ackley}, C.~{Adams}, T.~{Adams}, P.~{Addesso}, R.X. {Adhikari}, et~al.,
  \apj \textbf{839}, 12 (2017).
\newblock \doi{10.3847/1538-4357/aa677f}

\bibitem{2015RPPh...78k6901T}
R.~{Turolla}, S.~{Zane}, A.L. {Watts}, Reports on Progress in Physics
  \textbf{78}(11), 116901 (2015).
\newblock \doi{10.1088/0034-4885/78/11/116901}

\bibitem{2005Natur.434.1098H}
K.~{Hurley}, S.E. {Boggs}, D.M. {Smith}, R.C. {Duncan}, R.~{Lin},
  A.~{Zoglauer}, S.~{Krucker}, G.~{Hurford}, H.~{Hudson}, C.~{Wigger},
  W.~{Hajdas}, C.~{Thompson}, I.~{Mitrofanov}, A.~{Sanin}, W.~{Boynton},
  C.~{Fellows}, A.~{von Kienlin}, G.~{Lichti}, A.~{Rau}, T.~{Cline}, \nat
  \textbf{434}, 1098 (2005).
\newblock \doi{10.1038/nature03519}

\bibitem{2017arXiv171106593T}
H.~{Tong}, C.~{Yu}, ArXiv e-prints  (2017)

\bibitem{2017ApJ...848L..13A}
B.P. {Abbott}, R.~{Abbott}, T.D. {Abbott}, F.~{Acernese}, K.~{Ackley},
  C.~{Adams}, T.~{Adams}, P.~{Addesso}, R.X. {Adhikari}, V.B. {Adya}, et~al.,
  \apjl \textbf{848}, L13 (2017).
\newblock \doi{10.3847/2041-8213/aa920c}

\bibitem{2017PhRvL.119p1101A}
B.P. {Abbott}, R.~{Abbott}, T.D. {Abbott}, F.~{Acernese}, K.~{Ackley},
  C.~{Adams}, T.~{Adams}, P.~{Addesso}, R.X. {Adhikari}, V.B. {Adya}, et~al.,
  Physical Review Letters \textbf{119}(16), 161101 (2017).
\newblock \doi{10.1103/PhysRevLett.119.161101}

\bibitem{2015JHEAp...7...73D}
P.~{D'Avanzo}, Journal of High Energy Astrophysics \textbf{7}, 73 (2015).
\newblock \doi{10.1016/j.jheap.2015.07.002}

\bibitem{2008PhLB..668....1K}
P.G. {Krastev}, B.A. {Li}, A.~{Worley}, Physics Letters B \textbf{668}, 1
  (2008).
\newblock \doi{10.1016/j.physletb.2008.07.105}

\bibitem{2016ApJ...831...35D}
J.C.N. {de Araujo}, J.G. {Coelho}, C.A. {Costa}, \apj \textbf{831}, 35 (2016).
\newblock \doi{10.3847/0004-637X/831/1/35}

\bibitem{2017EPJC...77..350D}
J.C.N. {de Araujo}, J.G. {Coelho}, C.A. {Costa}, European Physical Journal C
  \textbf{77}, 350 (2017).
\newblock \doi{10.1140/epjc/s10052-017-4925-3}

\bibitem{2011MNRAS.412.1394L}
S.K. {Lander}, D.I. {Jones}, \mnras \textbf{412}, 1394 (2011).
\newblock \doi{10.1111/j.1365-2966.2010.17998.x}

\bibitem{2011MNRAS.412.1730L}
S.K. {Lander}, D.I. {Jones}, \mnras \textbf{412}, 1730 (2011).
\newblock \doi{10.1111/j.1365-2966.2010.18009.x}

\end{thebibliography}

\end{document}